**Self-Consistent Field Theory of Inhomogeneous Polymeric Systems: A Variational Derivation**

An-Chang Shi

Department of Physics and Astronomy, McMaster University

Hamilton, Ontario Canada L8S 4M1

**Abstract**

The self-consistent field theory (SCFT) is a powerful framework for the study of the phase behavior and structural properties of many-body systems. In particular, polymeric SCFT has been successfully applied to inhomogeneous polymeric systems such as polymer blends and block copolymer melts. The polymeric SCFT is commonly derived using field-theoretical techniques. Here we provide an alternative derivation of the SCFT equations and SCFT free energy functional using a variational principle. Numerical methods of solving the SCFT equations and applications of the SCFT are also briefly introduced.

**1. Introduction**

Inhomogeneous polymeric systems are ubiquitous in polymeric materials for various applications. Typical inhomogeneous polymeric materials are in the form of polymer blends, polymer solutions and block copolymers. Examples of the inhomogeneities found in these polymeric materials are the interfaces between different polymers found in polymer blends and the self-assembled microphases found in block copolymers [1]. Because of their rich phase behaviour and abundant technological applications, the study of inhomogeneous polymeric systems has attracted tremendous attention. Various theoretical frameworks have been developed to study the formation and structural properties of inhomogeneous polymeric materials. The ultimate goal of the theoretical study is to use the molecular properties of the polymers to predict thermodynamically stable phases and their phase transition boundaries, as well as the structural properties of these phases.

Among the various theoretical frameworks developed for the study of inhomogeneous polymeric systems, it could be argued that the most successful one is the self-consistent field theory (SCFT). From a fundamental perspective, the SCFT is a mean-field theory based on the statistical mechanics of many-body systems. For polymeric systems, the SCFT framework was initially developed by Edwards in 1960s [2]. Explicit application of the SCFT framework to the study of polymer blends and block copolymers was carried out by Helfand in 1975 [3]. Subsequent contributions to the development of polymeric SCFT were made by various authors and noticeably by Hong and Noolandi [4]. Since 1990s, applications of the SCFT to a large number of inhomogeneous polymeric systems have been made, resulting in a rich body of literature, including several books and review articles [5,6], on the development and application of polymeric SCFT.

The most successful application of the polymeric SCFT is found in the study of the phase behaviour of block copolymers. This success stems from the development of efficient and accurate numerical techniques over the last decades to solve the SCFT equations. Early attempts to obtain numerical solutions of the SCFT equations for block copolymers were made by a number of researchers including Helfand and Wasserman [7], Shull [8] and Whitmore and coworkers [9]. These authors constructed phase diagrams of block copolymers using various approximate numerical techniques. An important development of the numerical methods for polymeric SCFT is made by Matsen and Schick [10], who proposed a spectral-method based on the crystalline symmetry of the ordered phases. This method allowed them to obtain accurate numerical solutions of the SCFT equations corresponding to the known ordered phases of diblock copolymers at that time. Since then, the spectral-method of Matsen and Schick has been successfully applied to various block copolymer systems [11]. Subsequent development of the polymeric SCFT included the theory of Gaussian fluctuations in ordered phases [12] and various numerical methods for solving the SCFT equations [13,14]. Based on the large number of previous studies [15], it could be stated definitively that the SCFT provides a powerful theoretical framework for the study of inhomogeneous polymeric systems.

In general, a classical statistical mechanical system is usually specified using a particle-based description, in which the partition function of the system is obtained as a sum over all the degrees of freedom (positions and momenta) of the particles. The essence of the SCFT is to map the particle-based description of a statistical mechanical system into a description based on a set of collective variables or fields. In the field-based description, the partition function of the system is specified as a functional integral over the fields. The mapping between the particle-based description and the field-based formulation can be carried out exactly using field-theoretical techniques [5,6]. However, exact evaluation the functional integral is in general a formidable task. Mean-field theory of the system is obtained when the functional integral is evaluated using a saddle-point approximation. This approximation leads to a set of mean-field equations, or SCFT equations, determining the optimum fields. The SCFT equations should be solved analytically or numerically. These mean-field solutions are then used to describe the structure and property of the inhomogeneous polymeric phases.

In most, if not all, of the literatures on the polymeric SCFT, the development of the theory relies on field-theoretical techniques involving the introduction of imaginary conjugate fields [5,6]. One exception to this observation is found in the lattice version of a SCFT developed by Scheutjens and Fleer [16]. In this approach the probability of a polymer configuration in a mean-field is obtained using a transfer matrix method, whereas the mean-field for the polymer is constructed by averaging over the interactions from the neighbouring lattice sites. The spirit of this mean-field approach is similar to the Bragg-Williams theory originally developed for a lattice model of binary alloys [17].

The field-theoretical approach is exact and general. However, an intuitive physical meaning of the complex conjugate fields is not transparent. Therefore, it is desirable and useful to develop alternative approaches for the derivation of the polymeric SCFT. Here we provide such an alternative approach to derive the self-consistent field theory. Specifically, we will derive the SCFT of polymers using the Gibbs-Bogoliubov-Feynman variational principle [17,18,19]. The theoretical development results in a variational, or SCFT, free energy functional of the system. Minimization of the SCFT free energy functional leads to a set of SCFT equations. The resulting SCFT framework can be used to study the phase behaviour and structural properties of inhomogeneous polymeric systems. Although a variational approach is used to derive the theory, the resulting SCFT has exactly the same structure as the one derived using the field-theoretical approach. We emphasize that the variational approach to derive the SCFT theory is a general theoretical technique applicable to any statistical mechanical systems. We hope that the pedagogical account of the theoretical development presented here will serve as a useful introduction for the readers to learn and apply the SCFT to different statistical mechanical systems. Furthermore, we have also refrained from a detailed review of the numerical methods and practical applications of the SCFT in the current paper. This choice of materials is made based on our aim of presenting the variational principle as a general approach to the development of mean-field theory and the existence of extensive literature on the numerical methods and applications of SCFT to block copolymers.

The starting point of the theoretical framework is the development of a molecular model for the polymeric systems. To be specific, we present a detailed derivation of the theory using an AB diblock copolymer melt as a model system. As a first step, we formulate the theory by establishing a general variational principle. This variational approach is then used as the starting point to formulate a general variational mean-field theory, also known as the $\text{Tr}[\rho \ln \rho]$ mean-field theory [17], that is applicable to any statistical mechanics systems. This variational mean-field theory is then used to derive the SCFT equations of AB diblock copolymers.

## 2. Variational Mean-Field Theory

### 2.1 The Gibbs-Bogoliubov-Feynman inequality

The goal of the variational mean-field theory is to derive a variational free energy functional for a given statistical mechanical system. The starting point of our derivation is the well-known mathematical inequality that, for any real number $x$, $e^x \geq 1 + x$. This inequality is the result of the fundamental property, *i.e.* the convexity, of the exponential function. For the case where $\phi$ is a real discrete or continuous random variable with an associated probability distribution $p(\phi)$, we can define the expectation value of any function of $\phi$, $f(\phi)$, by the operation,

$$\langle f(\phi) \rangle = \text{Tr}[p(\phi) f(\phi)],$$

where the trace operator (Tr) indicates a sum or an integral over all possible values of the random variable $\phi$. Using the inequality $e^x \geq 1 + x$, it is straightforward to proof an inequality involving the expectation values,

$$\langle e^{-\lambda f(\phi)} \rangle = e^{-\lambda \langle f(\phi) \rangle} \langle e^{-\lambda (f(\phi) - \langle f(\phi) \rangle)} \rangle \geq e^{-\lambda \langle f(\phi) \rangle} \langle 1 - \lambda (f(\phi) - \langle f(\phi) \rangle) \rangle = e^{-\lambda \langle f(\phi) \rangle}.$$

This mathematical inequality, $\langle e^{-\lambda f(\phi)} \rangle \geq e^{-\lambda \langle f(\phi) \rangle}$, can be used to establish a variational formulation of statistical mechanics [17,18,19], as detailed below.

Consider a statistical mechanical system described by a classical Hamiltonian $\mathcal{H}[\phi]$, where $\phi$ is a discrete or continuous classical field. The canonical partition function of the system is given by,

$$Z = e^{-\beta F} = \text{Tr}\big[e^{-\beta \mathcal{H}[\phi]}\big],$$

where $\beta = 1/k_B T$ and $F$ is the Helmholtz free energy of the system associated with the Hamiltonian $\mathcal{H}[\phi]$. Let us now introduce an arbitrary normalized classical probability distribution, $\rho(\phi)$, which satisfies $\text{Tr}[\rho(\phi)] = 1$ and $\rho(\phi) \geq 0$. Using the identity $1 = \rho/\rho = \rho e^{-\ln \rho}$, we can write the partition function of the system in the form,

$$Z = \text{Tr}\big[\rho e^{-\beta \mathcal{H}[\phi] - \ln \rho}\big] = \langle e^{-\beta \mathcal{H}[\phi] - \ln \rho} \rangle_\rho,$$

where $\langle \cdots \rangle_\rho$ indicates the expectation value with respect to the distribution $\rho(\phi)$. Using the inequality, $\langle e^{-\lambda f(\phi)} \rangle \geq e^{-\lambda \langle f(\phi) \rangle}$, and the definition of the free energy, we obtain,

$$e^{-\beta F} = \langle e^{-\beta \mathcal{H}[\phi] - \ln \rho} \rangle_\rho \geq e^{-\beta \langle \mathcal{H}[\phi] \rangle_\rho - \langle \ln \rho \rangle_\rho}.$$

After some simple manipulations, this inequality leads to an inequality for the Helmholtz free energy, known-as the Gibbs-Bogoliubov-Feynman (GBF) inequality [17,18,19],

$$F \leq F_\rho = \langle \mathcal{H}[\phi] \rangle_\rho + k_B T \langle \ln \rho \rangle_\rho = \text{Tr}\big[\rho \mathcal{H}[\phi]\big] + k_B T\, \text{Tr}[\rho \ln \rho],$$

where $F_\rho$ is a variational free energy functional associated with the arbitrary and unknown probability distribution $\rho$. This inequality is completely general and valid for any probability distribution $\rho$. It is interesting to note that the variational free energy can be written in a very suggestive form,

$$F_\rho = \langle \mathcal{H}[\phi] \rangle_\rho - TS_\rho,$$

where the two terms, $\langle \mathcal{H}[\phi] \rangle_\rho$ and $S_\rho = -k_B \operatorname{Tr}[\rho \ln \rho]$, could be interpreted as the energetic and entropic contributions to the trial free energy functional.

The GBF inequality provides a variational principle to obtain the best probability distribution for a statistical mechanics system. In particular, minimizing $F_\rho$ with respect to the probability distribution $\rho$ subject to the constraint $\operatorname{Tr}[\rho(\phi)] = 1$ would result in the actual equilibrium probability distribution $\rho = \frac{e^{-\beta \mathcal{H}[\phi]}}{Z}$, i.e. the Boltzmann distribution. Therefore, the GBF variational principle does lead to the correct results for the system in this case.

**2.2 Mean-Field Theory via the GBF Variational Principle**

The Gibbs-Bogoliubov-Feynman inequality provides a generic approach to calculate the free energy of a statistical mechanical system using variational approximation [17,18,19]. The procedure starts with choosing a trial probability distribution $\rho$ containing certain free parameters to approximate the actual probability distribution. The best approximation within the scope of the trial probability distribution is then obtained by minimizing the variational free energy $F_\rho$ with respect to the parameters in $\rho$. There are many choices of the trial probability distributions, resulting in different approximations of the theory.

Mean-field theory is obtained by choosing a trial probability distribution $\rho$ that is written as the product of single-particle probability distributions. Let $\rho_i$ be a single-particle probability distribution depending only on the degree of freedom of the $i$-th particle, we can write the mean-field probability distribution as a product of the single-particle distributions,

$$\rho = \prod_i \rho_i.$$

Inserting this form of the probability distribution into the variational free energy functional, we obtain the variational mean-field free energy,

$$F_\rho = \langle \mathcal{H}[\phi] \rangle_\rho + k_B T \sum_i \text{Tr}[\rho_i \ln \rho_i].$$

The precise form of the energy contribution, $\langle \mathcal{H}[\phi] \rangle_\rho$, depends on the specific form of the Hamiltonian of the system, whereas the entropic term becomes a sum of contributions from single particles.

A number of approaches could be used to determine the variational minima of the mean-field free energy functional [17]. In what follows we will adopt the method of introducing a conjugate field, or the mean-field, $\omega_i$, and write the normalized single-particle probability distribution in terms of the mean-field as,

$$\rho_i = \frac{1}{Q_i} e^{-\omega_i \phi_i}$$

where $Q_i$ is a normalization constant. Specifically, this normalization constant is given by,

$$Q_i = \text{Tr}\left[e^{-\omega_i \phi_i}\right].$$

From this expression, it is clear that we can interpret $Q_i$ as the partition function of a single-particle in the field $\omega_i$. It is by design that this form of $\rho_i$ satisfies the normalization condition $\text{Tr}[\rho_i] = 1$. Furthermore, the expectation value of the field or the order parameter of the system within the mean-field approximation is given by,

$$\langle \phi_i \rangle = \text{Tr}[\rho_i \phi_i].$$

With this choice of the single-particle probability distribution, we can write the mean-field free energy functional in the form,

$$F_\rho = \langle \mathcal{H}[\phi] \rangle_\rho - k_B T \sum_i \omega_i \langle \phi_i \rangle - k_B T \sum_i \ln Q_i.$$

Minimizing this mean-field free energy functional with respect to the mean-fields $\omega_i$ and order parameter, $\langle\phi_i\rangle$, leads to a set of self-consistent mean-field equations,

$$\omega_i = \frac{\delta(\beta\langle\mathcal{H}[\phi]\rangle_\rho)}{\delta\langle\phi_i\rangle}$$
$$\langle\phi_i\rangle = -\frac{1}{Q_i}\frac{\delta Q_i}{\delta\omega_i}.$$

These two coupled equations are to be solved self-consistently to yield mean-field solutions. The mean-field free energy of the system is then obtained by inserting the mean-field solutions into the variational mean-field free energy. Different solutions of the mean-field equations correspond to different phases of the system. A comparison of the free energy of the different phases can be used to construct a phase diagram of the system. It is interesting to note that this set of mean-field equations has the same structure as the polymeric SCFT equations given below.

For many cases the Hamiltonian of a many-body system can be written as a sum of two terms,

$$\mathcal{H}[\phi] = \mathcal{H}_0[\phi] + \mathcal{V}[\phi],$$

where $\mathcal{H}_0[\phi] = \sum_i \mathcal{H}_{0,i}[\phi_i]$ and $\mathcal{H}_{0,i}[\phi_i]$ depends only on the single-particle degree of freedom of a single particle. $\mathcal{V}[\phi] = k_B T \mathcal{W}[\phi]$ is the contribution from the interactions between the particles. In this case it is convenient to include the non-interacting Hamiltonian, $\mathcal{H}_{0,i}[\phi_i]$, in the variational probability distribution,

$$\rho_i = \frac{1}{Q_i} e^{-\beta\mathcal{H}_{0,i}[\phi_i]-\omega_i\phi_i}.$$

With this, the single-particle partition function includes the contribution from the non-interacting single-particle Hamiltonian,

$$Q_i = \text{Tr}\left[e^{-\beta\mathcal{H}_{0,i}[\phi_i]-\omega_i\phi_i}\right].$$

Again, the quantity $Q_i$ can be interpreted as the partition function of one particle in the external field $\omega_i$. For this choice of the variational probability distribution, the mean-field free energy becomes,

$$\frac{F_\rho}{k_B T} = \langle \mathcal{W}[\phi] \rangle - \sum_i \omega_i \langle \phi_i \rangle - \sum_i \ln Q_i.$$

The mean-field equations of this system are obtained by minimizing the free energy functional with respect to $\langle \phi_i \rangle$ and $\omega_i$, resulting in the following self-consistent equations,

$$\begin{aligned} \omega_i &= \frac{\delta \langle \mathcal{W}[\phi] \rangle}{\delta \langle \phi_i \rangle} \\ \langle \phi_i \rangle &= -\frac{1}{Q_i} \frac{\delta Q_i}{\delta \omega_i} \end{aligned}.$$

This formulation of the mean-field theory using the Gibbs-Bogoliubov-Feynman variational principle provides a flexible framework to derive a mean-field theory for any statistical mechanical systems. The structure of the resulting mean-field equations resembles that of the polymeric SCFT equations [5,6]. In what follows we will illustrate the approach by applying the theory to a lattice gas model and to a model of AB diblock copolymers.

**2.2 Mean-Field Theory of a Lattice Gas**

The application of the Gibbs-Bogoliubov-Feynman variational principle is nicely illustrated by the derivation of a mean-field free energy functional for the lattice gas model. The lattice gas model is a statistical mechanical model for the study of liquid-vapor phase transitions [20]. Within the lattice gas model, the particles are living on a lattice. The $i$-th lattice site is assigned an occupation variable $n_i$, where $n_i = 0$ and $n_i = 1$ indicates empty and occupied sites, respectively. The hardcore interaction between the particles is modeled by the constraint that each lattice site can only be empty or occupied by at the most one particle. The interactions between the particles are assumed to be described by a pairwise interaction potential,

$$V_{ij} = V(r_i - r_j).$$

With these assumptions, the Hamiltonian of a lattice gas is written as,

$$\mathcal{H}[n_i] = \frac{1}{2}\sum_{i,j} n_i V_{ij} n_j - \sum_i \mu n_i,$$

where $\mu$ is the chemical potential that controls the density of the particles in the system. It is interesting to note that the lattice gas model can be transformed to the Ising model of magnetism via the transformation $s_i = 2n_i - 1 = \pm 1$ [20].

For this simple lattice gas model, the single-particle degree of freedom is the two states corresponding to $n_i = 0, 1$. In this case the single-particle probability distribution can be written in the form,

$$\rho_i = \frac{1}{Q_i} e^{-\omega_i n_i},$$

where the single-particle partition function is easily found as,

$$Q_i = \sum_{n_i=0,1} e^{-\omega_i n_i} = 1 + e^{-\omega_i}.$$

Here the average density of the particles is given by,

$$\phi_i = \langle n_i \rangle = \frac{1}{Q_i} \sum_{n_i=0,1} n_i e^{-\omega_i n_i} = \frac{e^{-\omega_i}}{1 + e^{-\omega_i}}.$$

Furthermore, the average energy of the system within the proposed variational probability distribution is given by,

$$\langle \mathcal{H}[n_i] \rangle = \frac{1}{2}\sum_{i,j} \phi_i V_{ij} \phi_j - \sum_i \mu \phi_i$$

With this form of the mean-field probability distribution, the mean-field equations become,

$$\omega_i = \frac{\delta(\beta\langle\mathcal{H}[\phi]\rangle_\rho)}{\delta\langle\phi_i\rangle} = \beta\left(\sum_j V_{ij}\phi_j - \mu\right).$$

$$\phi_i = -\frac{1}{Q_i}\frac{\delta Q_i}{\delta\omega_i} = \frac{e^{-\omega_i}}{1+e^{-\omega_i}}$$

This set of mean-field equations are in the form of a self-consistent field theory (SCFT). Specifically, the two variables, $\phi_i$ and $\omega_i$, are determined self-consistently from the coupled mean-field equations. The structure of the theory is similar to the structure of the polymeric SCFT.

In the literature, the mean-field theory of lattice gas is commonly cast in the form of a density functional theory. This can be achieved by solving the $\omega_i$ field as a function of the density $\phi_i$. Because of the simplicity of the lattice gas model, we can obtain an explicit expression of the mean-field, $\omega_i$, in terms of the order parameter, $\phi_i$,

$$\omega_i = \ln\frac{1-\phi_i}{\phi_i}.$$

Using these relations to eliminate $\omega_i$ from the mean-field free energy, we obtain an expression of the mean-field free energy in terms of the order parameter or density $\phi_i$ alone,

$$F_\rho = \frac{1}{2}\sum_{i,j}\phi_i V_{ij}\phi_j + k_B T \sum_i [\phi_i \ln \phi_i + (1-\phi_i)\ln(1-\phi_i)] - \sum_i \mu\phi_i.$$

This expression of the free energy in terms of the density profile has the standard form of a density functional theory derived by other approaches [19]. The mean-field equations of the system are obtained by minimizing the free energy with respect to the density $\phi_i$,

$$\frac{\delta F_\rho}{\delta \phi_i} = \sum_j V_{ij}\phi_j + k_B T \ln \frac{\phi_i}{1-\phi_i} - \mu = 0.$$

This mean-field equation is equivalent to the SCFT equations given above. The solutions of this mean-field equation correspond to the liquid and vapor phases of the system. The phase diagram of the system can also be constructed by the mean-field solutions.

## 3. Variational Derivation of the SCFT for AB Diblock Copolymers

In this section we present a detailed derivation of the mean-field theory or the self-consistent field theory (SCFT) for inhomogeneous polymeric systems using the Gibbs-Bogoliubov-Feynman variational principle. We will use an AB diblock copolymer melt as a model system in our derivation. Extension of this variational approach to other systems with different polymeric architectures is straightforward. As the first step of developing the SCFT for polymeric systems, we specify a molecular model for the polymers, in which the polymers are modeled as Gaussian chains. Again, it is straightforward to extend the variational derivation to other chain models such as semiflexible chains or discrete chains. We then design a probability distribution of the model system by introducing a set of mean-fields. This variational probability distribution is used to derive a variational free energy functional for the block copolymers that has exactly the same form as the SCFT free energy derived using the field-theoretical techniques. Minimization of the free energy functional leads to the SCFT equations of the system. We emphasize that, although we obtain the same SCFT free energy functional as it should be, the variational derivation of the SCFT is physically more intuitive. Furthermore, the variational derivation does not involve complex variables, thus avoiding the complexity of the field-theoretical approach.

### 3.1 Molecular Model of AB Diblock Copolymers

We consider a model system of AB diblock copolymers, which are macromolecules composed of two sub-chains or blocks, labeled as *A*-block and *B*-block, respectively, tethered together at their ends. We assume that our system consists of $n_c$ diblock copolymer chains contained in a volume $V$. The A- and B-block of a diblock copolymer consists of $N_\alpha$ monomers of species $\alpha = A$ or $\alpha = B$. The total degree of polymerization or length of the diblock copolymer chain is therefore given by $N = N_A + N_B$. The volume or length fraction of the A-block and B-block is $f = \frac{N_A}{N}$ and $1 - f = \frac{N_B}{N}$, respectively. Furthermore, each segment is assumed to be associated with a Kuhn length $b_\alpha = \sigma_\alpha b$ ($\alpha = A, B$), where $b$ is a reference Kuhn length and $\sigma_\alpha$ characterizes the conformational asymmetry between the A- and B-blocks. For simplicity, the monomers are assumed to have the same monomer density, $\rho_0$, defined as monomers per unit volume. Thus, the hardcore volume per monomer is $\rho_0^{-1}$. In the final expressions we will use the convention that all lengths are scaled by the Gaussian radius of gyration of the diblock copolymers, $R_g = b\sqrt{N/6}$. Furthermore, the degree of polymerization of the diblock copolymers $N$ is used as the scale of the chain arc length. We also notice that the incompressibility condition of the system implies that $\frac{n_c N}{\rho_0 V} = 1$, thus the number of polymer chains and the volume of the system are not independent parameters of the model system.

A polymer chain is modeled as a continuous space curve. A function $v(s)$ is used to specify the architecture of the diblock copolymers, where $s$ labels the arc-length of the polymer chain. Specifically, an AB diblock copolymer is specified by labelling the type of monomers on the chain using the following function $v(s)$,

$$v(s) = \begin{cases} A & \text{if } 0 < s < N_A \\ B & \text{if } N_A < s < N \end{cases}.$$

Obviously other types of block copolymers would be specified by different forms of the function $v(s)$. The microscopic state, or the configuration, of a diblock copolymer is described by a space curve $\vec{R}_{v(s),i}(s)$, which is a function specifying the position of the $s$-th monomer of the $i$-th chain of type $v(s) = A, B$. For a given set of chain configurations, $\{\vec{R}_{v(s),i}(s)\}$, the concentrations, or density profiles, of the A and B monomers at a given spatial position $\vec{r}$ are given by,

$$\hat{\phi}_A(\vec{r}) = \frac{1}{\rho_0} \sum_{i=1}^{n_c} \int_0^{N_A} ds\, \delta[\vec{r} - \vec{R}_{v(s),i}(s)],$$

$$\hat{\phi}_B(\vec{r}) = \frac{1}{\rho_0} \sum_{i=1}^{n_c} \int_{N_A}^{N} ds\, \delta[\vec{r} - \vec{R}_{v(s),i}(s)],$$

where the hat on $\hat{\phi}_\alpha(\vec{r})$ indicates the fact that these concentrations are a functional of the chain configurations $\{\vec{R}_{v(s),i}(s)\}$. The average values of these concentration fields of the polymers, $\langle\hat{\phi}_\alpha(\vec{r})\rangle$, are taken as the order parameter of our system. In particular, an inhomogeneous phase of the system is described by a non-uniform average density profile, whereas an ordered crystalline phase of the system is described by an average density profile that is a periodic function in space. The specific form of the periodic function is determined by the symmetry of the ordered phase.

For simplicity, we will assume that the polymer chains are flexible Gaussian chains. It should be noted that other polymeric models such as semiflexible chains or helical wormlike chains could be used as well. For a Gaussian chain, the probability distribution $p_0[\vec{R}_{\alpha,i}(s)]$ is given by the standard Edwards model or the Wiener form [5,6],

$$p_0[\vec{R}_{\alpha,i}(s)] = A_\alpha \exp\left[-\frac{3}{2b_\alpha^2}\int_0^{N_\alpha} ds \left(\frac{d\vec{R}_{\alpha,i}(s)}{ds}\right)^2\right],$$

where $A_\alpha$ is a normalization constant. Because a diblock copolymer is composed of A- and B-blocks connected at their ends, the probability distribution, $P_0(\{\vec{R}(s)\})$, of a diblock copolymer found in a given configuration, $\{\vec{R}_{\alpha,i}(s)\}$, can be constructed by connecting the probability distributions of individual blocks or sub-chains,

$$P_0[\vec{R}_{\alpha,i}(s)] = p_0[\vec{R}_{A,i}(s)]p_0[\vec{R}_{B,i}(s)]\delta\left(\vec{R}_{A,i}(N_A) - \vec{R}_{B,i}(N_A)\right),$$

where a delta-function is introduced to ensure the connective of the diblock copolymers at the ends of the A and B blocks. Alternative, we can use the function, $v(s)$, to write the probability of a diblock copolymer chain in a compact form,

$$P_0[\vec{R}_{v(s),i}(s)] \equiv e^{-\beta \mathcal{H}_{0,i}} = A \exp\left[-\int_0^N ds\, \frac{3}{2b_{v(s)}^2}\left(\frac{d\vec{R}_{v(s),i}(s)}{ds}\right)^2\right],$$

where $A$ is a normalization constant and we have introduced the single-chain Hamiltonian $\mathcal{H}_{0,i}$ to describe the bonding interaction of the $i$-th block copolymer chain. The single-chain Hamiltonian is defined such that the single chain probability distribution is normalized,

$$\text{Tr}[e^{-\beta \mathcal{H}_{0,i}}] = \text{Tr}\left[P_0[\vec{R}_{v(s),i}(s)]\right] = \int \mathcal{D}\{\vec{R}_{\alpha,i}(s)\}\, P_0[\vec{R}_{\alpha,i}(s)] = 1.$$

It should be noted that the above expressions for the single-chain Hamiltonian are specific for AB diblock copolymers. For other more complex block copolymers such as ABC linear triblock copolymers and ABC star-block copolymers, these expressions should be modified such that the the sun-chains or blocks are connected according to the architecture of the copolymers.

### 3.2 Variational Derivation of the SCFT of Diblock Copolymers

We can now apply the GBF variational principle described above to derive a mean-field theory, corresponding to the self-consistent field theory, of diblock copolymers. For a system composed of $n_c$ diblock copolymer chains contained in a volume $V$, the Hamiltonian is given by,

$$\mathcal{H}[\phi] = \mathcal{H}_0[\phi] + \mathcal{V}[\phi],$$

where the ideal Hamiltonian, $\mathcal{H}_0[\phi]$, is written as a sum of single-chain contributions,

$$\mathcal{H}_0[\phi] = \sum_i \mathcal{H}_{0,i}.$$

The single-chain Hamiltonian, $\mathcal{H}_{0,i}$, describes the contribution from the bonding interactions of the polymer. The specific form of the single-chain Hamiltonian depends on the nature of the polymers. For a diblock copolymer modeled as a Gaussian chain, the single-chain Hamiltonian is defined via the Edwards model,

$$e^{-\beta \mathcal{H}_{0,i}} = P_0[\vec{R}_{v(s),i}(s)] = A \exp\left[-\int_0^N ds \, \frac{3}{2b_{v(s)}^2}\left(\frac{d\vec{R}_{v(s),i}(s)}{ds}\right)^2\right].$$

The interaction contribution to the Hamiltonian, $\mathcal{V}[\phi]$, is due to the intermolecular interactions between the different monomers. Following standard approach, we assume the interactions are short-ranged such that the interaction potential has the standard Flory-Huggins form [6],

$$\mathcal{V}[\phi] = k_B T \mathcal{W}(\{\hat{\phi}\}) = \rho_0 k_B T \chi \int d\vec{r} \, \hat{\phi}_A(\vec{r}) \hat{\phi}_B(\vec{r}),$$

where $\chi$ is the Flory-Huggins parameter which varies with temperature. Furthermore, we assume that the system is incompressible, approximating the hardcore interactions between the monomers. A Lagrange multiplier will be introduced later to enforce the incompressibility.

In order to apply the variational mean-field principle to the diblock copolymer system, we first introduce a variational probability distribution function for the whole system,

$$\rho = \frac{1}{Q} \exp\left[-\beta \mathcal{H}_0[\phi] - \int dr \sum_{\alpha=A,B} \rho_0 \omega_\alpha(r) \hat{\phi}_\alpha(r)\right],$$

where $Q = \text{Tr}[\rho]$ is a normalization factor and the factor $\rho_0$ is introduced for later convenience. The average concentrations or density profiles, $\langle \hat{\phi}_\alpha(\vec{r}) \rangle$, of the polymers associated with this variational probability distribution is given by,

$$\phi_\alpha(\vec{r}) \equiv \langle \hat{\phi}_\alpha(\vec{r}) \rangle = \text{Tr}[\rho \hat{\phi}_\alpha(\vec{r})] = -\frac{1}{\rho_0} \frac{\delta \ln Q}{\delta \omega_\alpha(\vec{r})}.$$

Because of the definition of the density operators $\hat{\phi}_\alpha(\vec{r})$, we can show that the variational probability distribution can be reduced to the form of a product of single-molecular probability distributions. Specifically, using the fact that the ideal Hamiltonian of the polymers is written as a sum of the contributions from each polymer, $\mathcal{H}_0[\phi] = \sum_i \mathcal{H}_{0,i}$, and the expression for the density operators $\hat{\phi}_\alpha(\vec{r})$, it is straightforward to write the total variational probability distribution as a product of the single-molecular partition function,

$$\rho = \prod_{i=1}^{n_c} \rho_i,$$

where the single-molecular variational probability distribution $\rho_i$ is given by,

$$\rho_i = \frac{1}{Q_c} \exp\left[-\int_0^N ds \left\{\frac{3}{2b_{v(s)}^2}\left(\frac{d\vec{R}_{v(s),i}(s)}{ds}\right)^2 + \omega_{v(s)}[\vec{R}_{v(s),i}(s)]\right\}\right].$$

Here the single-molecular partition function, $Q_c$, is obtained by adding contributions from all possible chain configurations. Specifically, $Q_c$ is given in the form of a path integral,

$$Q_c = \int \mathcal{D}\{\vec{R}_{\alpha,i}(s)\} \exp\left[-\int_0^N ds \left\{\frac{3}{2b_{v(s)}^2}\left(\frac{d\vec{R}_{v(s),i}(s)}{ds}\right)^2 + \omega_{v(s)}[\vec{R}_{v(s),i}(s)]\right\}\right].$$

From this expression, it is obvious that the single-chain partition function, $Q_c$, is a functional of the mean-fields, $\omega_\alpha(\vec{r})$, that is, $Q_c = Q_c(\{\omega_\alpha\})$. Because the chain configurations, $\vec{R}_{\alpha,i}(s)$, are summed over in this expression, the single-molecular partition function is the same for all the diblock copolymers. Therefore, the total mean-filed partition function becomes,

$$Q = \prod_{i=1}^{n_c} Q_c = (Q_c)^{n_c}.$$

The average concentration of the A- and B-blocks is now given in terms of functional derivatives of the single-chain partition function $Q_c$,

$$\phi_\alpha(\vec{r}) \equiv \langle\hat{\phi}_\alpha(\vec{r})\rangle = -\frac{n_c}{\rho_0}\frac{\delta \ln Q_c}{\delta \omega_\alpha(\vec{r})}.$$

With this choice of the variational probability distribution, the average energy of the system is given in terms of the polymer density profiles,

$$\langle \mathcal{H}[\phi] \rangle = \langle \mathcal{H}_0[\phi] \rangle + \langle \mathcal{V}[\phi] \rangle = \langle \mathcal{H}_0[\phi] \rangle + \rho_0 k_B T \chi \int d\vec{r}\, \phi_A(\vec{r})\phi_B(\vec{r}).$$

Similarly, the entropic contribution to the free energy functional of the system has the form,

$$-TS_\rho = k_B T \langle \ln \rho \rangle = -\langle \mathcal{H}_0[\phi] \rangle - k_B T \sum_\alpha \int dr\, \rho_0 \omega_\alpha(r) \phi_\alpha(r) - n_c k_B T \ln Q_c.$$

Assembling the expressions of the energy and entropy together, we obtain a variational, or SCFT, free energy functional of a diblock copolymer melt,

$$\frac{F_\rho}{\rho_0 k_B T} = \int d\vec{r} \left[ \chi \phi_A(\vec{r})\phi_B(\vec{r}) - \sum_\alpha \omega_\alpha(\vec{r}) \phi_\alpha(r) \right] - \frac{V}{N} \ln Q_c,$$

where we have used the incompressibility condition to relate the number of chains with the volume of the system, i.e. $\frac{n_c N}{\rho_0 V} = 1$. This variational mean-field free energy functional derived from the variational principle is exactly the same as the one derived using the field-theoretical techniques. Therefore, the variational mean-field approach to the SCFT is equivalent to the saddle-point approximation in the field-theoretical approach.

The mean-field equations, or the SCFT equations, are obtained by minimizing the variational free energy functional with respect to the densities $\phi_\alpha(\vec{r})$ and the mean-fields $\omega_\alpha(\vec{r})$. Minimization the free energy functional with respect to $\phi_\alpha(\vec{r})$ results in the first set of SCFT equations,

$$\omega_A(\vec{r}) = \chi \phi_B(\vec{r}) + \eta(\vec{r}),$$
$$\omega_B(\vec{r}) = \chi \phi_A(\vec{r}) + \eta(\vec{r}),$$

where the function $\eta(\vec{r})$ is a Lagrange multiplier introduced to enforce the incompressibility condition of the system,

$$\phi_A(\vec{r}) + \phi_B(\vec{r}) = 1.$$

The second set of the SCFT equations is obtained by minimizing the free energy functional with respect to the fields $\omega_\alpha(\vec{r})$,

$$\phi_\alpha(\vec{r}) = -\frac{V}{N}\frac{\delta \ln Q_c}{\delta \omega_\alpha(\vec{r})}.$$

Explicit expression of the density distributions can be obtained in terms of the chain propagators as described below.

### 3.3 The Single-Chain Partition Function and Chain Propagators

The SCFT free energy and density distributions are given in terms of the single-molecule, or single-chain, partition function, $Q_c = Q_c(\{\omega_\alpha\})$, which is the partition of one diblock copolymer chain in the fields $\omega_\alpha(\vec{r})$. In order to proceed, it is helpful to sepcify some properties of the single-chain partition function [5,6].

The single-chain partition function, $Q_c$, is obtained by summing over the contributions from all the chain configurations in the presence of an external field $\omega_\alpha(\vec{r})$. Explicitly, the single-chain partition function is specified as a path-integral,

$$Q_c = \int \mathcal{D}\{\vec{R}(s)\} \exp\left[-\int_0^N ds \left\{\frac{3}{2b_{v(s)}^2}\left(\frac{d\vec{R}(s)}{ds}\right)^2 + \omega_{v(s)}[\vec{R}(s)]\right\}\right],$$

where we have dropped the sub-index of the chain configuration, i.e. $\vec{R}_{\alpha,i}(s) \to \vec{R}(s)$, because the path integral is the same for all the chains. It should be emphasized that expressing the single-chain partition function as a path integral is an elegant formulation. In particular, this formulation resembles the Feynman's path integral of quantum mechanics. As such, many concepts and techniques developed in quantum mechanics could be applied to the statistical mechanics of polymers.

For computational purpose, however, it is useful to introduce chain propagators, or Green functions, $Q_\alpha(\vec{r}, s|\vec{r}')$, of the polymers, such that the single-chain partition function $Q_c$ can be expressed in terms of these functions,

$$Q_c(\{\omega_\alpha\}) = \int d\vec{r}_1 d\vec{r}_2 d\vec{r}_3 \ Q_B(\vec{r}_1, N_B|\vec{r}_2) Q_A(\vec{r}_2, N_A|\vec{r}_3).$$

Here the chain propagators are defined as a constrained path integral over the chain configurations,

$$Q_\alpha(\vec{r}, s|\vec{r}') = \int_{\vec{R}(0)=\vec{r}'}^{\vec{R}(s)=\vec{r}} \mathcal{D}\{\vec{R}(s)\} e^{-\int_0^{N_\alpha} ds \left[\frac{3}{2b_\alpha^2}\int_0^{N_\alpha} ds \left(\frac{d\vec{R}(s)}{ds}\right)^2 + \omega_\alpha(\vec{R}(s))\right]}.$$

The physical interpretation of the propagators is that the quantity, $Q_\alpha(\vec{r}, s|\vec{r}')$, specifies the conditional probability distribution of finding the $s^{\text{th}}$-segment of the polymer at position $\vec{r}$, given that the $0^{\text{th}}$-segment of the polymer is at position $\vec{r}'$, in the presence of an external field $\omega_\alpha(\vec{r})$. A direct evaluation of the path integrals is usually a difficult task. Fortunately, there is an alternative and more convenient method to calculate the propagator. This method is based on the fact that the propagators are solutions of a differential equation, commonly referred as the modified diffusion equation [6],

$$\frac{\partial}{\partial s} Q_\alpha(\vec{r}, s|\vec{r}') = \frac{b_\alpha^2}{6} \nabla^2 Q_\alpha(\vec{r}, s|\vec{r}') - \omega_\alpha(\vec{r}) Q_\alpha(\vec{r}, s|\vec{r}'),$$

with the initial conditions $Q_\alpha(\vec{r}, 0|\vec{r}') = \delta(\vec{r} - \vec{r}')$. Therefore, instead of performing integrations over the paths, these propagators can be obtained by solving the modified diffusion equation. It is interesting to note that the modified diffusion equation could be viewed as the Schrödinger equation with imaginary time. Therefore, the transformation from the path integral description to the modified diffusion equation is equivalent to the transformation from the Feynman's path integral formulation of quantum mechanics to the Schrödinger equation.

In many applications of the SCFT, especially in the computation of the density fields, it is useful to introduce end-integrated propagators, $q_\alpha(\vec{r}, s)$, defined by,

$$q_\alpha(\vec{r}, s) = \int d\vec{r}' Q_\alpha(\vec{r}, s|\vec{r}').$$

The end-integrated propagators satisfy the same partial differential equation as $Q_\alpha(\vec{r}, s|\vec{r}')$, but with the initial condition, $q_\alpha(\vec{r}, 0) = 1$. Physically, $q_\alpha(\vec{r}, s)$ represents the probability distribution of the $s^{\text{th}}$-segment at position $\vec{r}$, irrespective to where the end-segment at $s = 0$ is, in the presence of an external field $\omega_\alpha(\vec{r})$.

A linear polymer chain of length $N$ has two ends at $s = 0$ and $s = N$. Thus a particular segment at $s$ is associated with two end-integrated propagators, $q_\alpha(\vec{r}, s)$ and $q_\alpha^\dagger(\vec{r}, s)$, reaching to that particular segment starting from one of these two ends. For a homopolymer chain, these two end-integrated propagators are identical because of the symmetry of the polymer chain. For more complex polymers such as block copolymers, however, $q_\alpha(\vec{r}, s)$ and $q_\alpha^\dagger(\vec{r}, s)$ are in general different because the polymers are in general not symmetric. For example, an AB diblock copolymer have two non-equivalent ends, thus the two propagators, $q_\alpha(\vec{r}, s)$ and $q_\alpha^\dagger(\vec{r}, s)$, for a diblock copolymer are different. These propagators must be computed separately. Specifically, the propagator $q_\alpha^\dagger(\vec{r}, s)$ satisfies the same modified diffusion equation but with different initial conditions, $q_A^\dagger(\vec{r}, 0) = q_B(\vec{r}, N_B)$, and $q_B^\dagger(\vec{r}, 0) = q_A(\vec{r}, N_A)$.

The single chain partition function is given in terms of the end-integrated propagators,,

$$Q_c = \int d\vec{r}\, q_A^\dagger(\vec{r}, N).$$

The derivation of the mean-field equations for the density distributions requires expressions of the functional derivatives of the single-chain partition functions with respect to the fields $\omega_\alpha(\vec{r})$. These functional derivatives are given by [5],

$$\frac{\delta Q_\alpha(\vec{r}_1, s|\vec{r}_2)}{\delta \omega_\alpha(\vec{r})} = -\int_0^s ds' Q_\alpha(\vec{r}_1, s - s'|\vec{r}) Q_\alpha(\vec{r}, s'|\vec{r}_2).$$

**3.4 Scaled Form of the SCFT Equations of Diblock Copolymers**

In the final expression of the SCFT equations, we will use the convention that all lengths are scaled by the Gaussian radius of gyration of the diblock copolymers, $R_g = b\sqrt{N/6}$. Furthermore, the arc length of the chain is scaled by the degree of polymerization of the diblock copolymers $N$. The derivation of the scaled form of the SCFT equations starts with writing the modified diffusion equation in the form,

$$N \frac{\partial}{\partial s} Q_\alpha(\vec{r}, s|\vec{r}') = \frac{Nb^2 \sigma_\alpha^2}{6} \nabla^2 Q_\alpha(\vec{r}, s|\vec{r}') - N\omega_\alpha(\vec{r}) Q_\alpha(\vec{r}, s|\vec{r}').$$

We will now scale all the lengths and arc-length using $R_g$ and $N$ as the scales. Furthermore, we scale the fields $\omega_\alpha(\vec{r})$ by adsorbing the factor $N$ into it, $N\omega_\alpha(\vec{r}) \to \omega_\alpha(\vec{r})$. With these changes, the modified diffusion equation assumes the scaled form given by,

$$\frac{\partial}{\partial s} Q_\alpha(\vec{r}, s|\vec{r}') = \sigma_\alpha^2 \nabla^2 Q_\alpha(\vec{r}, s|\vec{r}') - \omega_\alpha(\vec{r}) Q_\alpha(\vec{r}, s|\vec{r}'),$$

with the initial conditions, $Q_\alpha(\vec{r}, 0|\vec{r}') = \delta(\vec{r} - \vec{r}')$. Here the arc-length variable changes from $s = 0$ to $s = f$ for the A-blocks and from $s = f$ to $s = 1$ for the B-blocks.

The end-integrated propagators, $q_\alpha(\vec{r}, s)$, are given by,

$$q_\alpha(\vec{r}, s) = \int d\vec{r}' Q_\alpha(\vec{r}, s|\vec{r}').$$

It is convenient to use the function $v(s)$ ($v(s) = A$ for $0 < s < f$ and $v(s) = B$ for $f < s < 1$) to label the Kuhn length and the fields. This allows us to write the end-integrated propagators using a single expression $q(\vec{r}, s)$, which satisfy the modified differential equation of the form,

$$\frac{\partial}{\partial s} q(\vec{r}, s) = \sigma_{v(s)}^2 \nabla^2 q(\vec{r}, s) - \omega_{v(s)}(\vec{r}) q(\vec{r}, s),$$

with the initial condition, $q(\vec{r}, 0) = 1$. Similarly, it is convenient to cast the propagator $q^\dagger(\vec{r}, s)$ as propagating from $s = 1$ to $s = 0$, thus it satisfies the modified diffusion equation,

$$-\frac{\partial}{\partial s} q^\dagger(\vec{r}, s) = \sigma_{v(s)}^2 \nabla^2 q^\dagger(\vec{r}, s) - \omega_{v(s)}(\vec{r}) q^\dagger(\vec{r}, s),$$

with a different initial condition, $q^\dagger(\vec{r}, 1) = 1$. The single-chain partition function can now be written in terms of these end-integrated propagators as,

$$Q_c = \frac{1}{V} \int d\vec{r}\, q(\vec{r}, s) q^\dagger(\vec{r}, s),$$

where the factor of $\frac{1}{V}$ is introduced for convenience.

Putting all pieces together, the SCFT free energy functional for a diblock copolymer melt in the scaled form is given by the expression,

$$\frac{F}{n_c k_B T} = \frac{1}{V} \int d\vec{r}\, [\chi N \phi_A(\vec{r}) \phi_B(\vec{r}) - \omega_A(\vec{r}) \phi_A(\vec{r}) - \omega_B(\vec{r}) \phi_B(\vec{r})] - \ln Q_c.$$

This expression gives the free energy per chain in unit of $k_B T$. The corresponding SCFT equations are obtained by minimizing the free energy with respect to $\phi_\alpha(\vec{r})$ and $\omega_\alpha(\vec{r})$ fields. Specifically, the set of SCFT equations are given by,

$$\phi_A(\vec{r}) = \frac{1}{Q_c} \int_0^f ds\, q(\vec{r},s)q^\dagger(\vec{r},s),$$

$$\phi_B(\vec{r}) = \frac{1}{Q_c} \int_f^1 ds\, q(\vec{r},s)q^\dagger(\vec{r},s),$$

$$\omega_A(\vec{r}) = \chi\phi_B(\vec{r}) + \eta(\vec{r}),$$

$$\omega_B(\vec{r}) = \chi\phi_A(\vec{r}) + \eta(\vec{r}),$$

$$\phi_A(\vec{r}) + \phi_B(\vec{r}) = 1.$$

Here the field $\eta(\vec{r})$ is a Lagrange multiplier to ensure the incompressibility condition.

For a diblock copolymer melt, the SCFT equations are composed of five equations with five unknown variables, $\{\phi_A(\vec{r}), \phi_B(\vec{r}), \omega_A(\vec{r}), \omega_B(\vec{r}), \eta(\vec{r})\}$. Solutions of the SCFT equations are local minima of the free energy functional, corresponding to stable and metastable phases of the system. Within the mean-field approximation, there are several parameters controlling the thermodynamic properties of the system, including the volume fraction of the block copolymer quantified by $f$, the relative Kuhn lengths $\sigma_\alpha$ of the blocks and the interaction strength between the A and B blocks quantified by the product $\chi N$.

## 4. Numerical Techniques for Solving SCFT Equations

For a given set of the control parameters, $\{f, \chi N, \sigma_A, \sigma_B\}$, for diblock copolymers, the SCFT equations can have different solutions, corresponding to different local minima of the SCFT free energy functional. Physically, these different local minima correspond to different phases of the systems. A phase diagram of diblock copolymers can be constructed by comparing the free energy of these different phases. Furthermore, the SCFT solutions contain information about the symmetry and structure of the ordered phases. However, it should be noted that the SCFT equations are a set of nonlinear and nonlocal equations. Therefore, finding solutions of the SCFT equations is a challenging task. Due to the efforts of a large number of researchers in the past decades, a number of efficient and accurate numerical methods to solve the SCFT equations for inhomogeneous polymeric systems have been developed [6,14]. In what follows, we give a brief review of the numerical methods for solving the SCFT equations of block copolymers.

From a computational perspective, finding solutions of the SCFT equations involves two major steps: (1) solving the modified diffusion equations to obtain the propagators, $\{q(\vec{r},s), q^\dagger(\vec{r},s)\}$, and compute the concentration fields, $\phi_\alpha(\vec{r})$, and (2) evolving the fields, $\{\phi_\alpha(\vec{r}), \omega_\alpha(\vec{r}), \eta(\vec{r})\}$, iteratively to achieve self-consistency. A generic algorithm, the so-called Picard-type algorithm, is often used to evolve the fields. This self-consistent iteration process proceeds as follows: (i) initializing the mean-fields $\omega_\alpha(\vec{r})$, (ii) solving the modified diffusion equations to obtain the propagators $\{q(\vec{r},s), q^\dagger(\vec{r},s)\}$, (iii) computing the single-chain partition function $Q_c$ and the concentration fields $\phi_\alpha(\vec{r})$, (iv) updating the mean-fields $\{\omega_\alpha(\vec{r}), \eta(\vec{r})\}$ using the new concentrations $\phi_\alpha(\vec{r})$. Because of the iterative nature of the algorithm, the solution of the computation depends on the initial configurations. For random initial fields, this iterative process could lead to non-trivial solutions in many cases. On the other hand, for a given initial field with specified symmetry, this iteration process often leads to a converged solution corresponding to an ordered phase of the system with the specified symmetry. The free energy of the different SCFT solutions determines the relative stability of these phases.

Computationally, the most demanding step of finding numerical solutions of the SCFT equations is to obtain the propagators $\{q(\vec{r},s), q^\dagger(\vec{r},s)\}$ for a given mean-fields $\omega_\alpha(\vec{r})$, which amounts to solve the modified diffusion equation with proper initial and boundary conditions. Depending on the methods of discretization, numerical methods to solve the modified diffusion equation have been developed in real-space and reciprocal-space. Details of implementing the real-space and reciprocal-space methods to solve the SCFT equations are found in several review articles and book chapters [15]. The real-space and reciprocal-space methods have their own advantages and drawbacks. Fruitful applications of these methods to solve the SCFT equations are abundant in the literature [6,10-15].

Besides the real-space and reciprocal-space methods, a hybrid method, or the pseudo-spectral method, utilizing computations in real-space and reciprocal-space simultaneously has been proposed to solve the modified diffusion equations [14]. The pseudo-spectral method is based on the observation that any function can be described in real-space or in Fourier space, and interchanges between these spaces can be carried out efficiently using fast Fourier transform packages. Computation tasks are to be performed in either real-space or reciprocal-space according to the structure of the functions. Combined with efficient iteration steps, the pseudo-spectral method provides a highly efficient and accurate algorithm to solve the SCFT equations.

The pseudo-spectral method to solve the modified diffusion equations is based on the observation that solution of the modified diffusion equation could be obtained using a time-evolution operation,

$$q(\vec{r}, s + ds) = \exp\left[ds\left(\sigma_{v(s)}^2 \nabla^2 - \omega_{v(s)}(\vec{r})\right)\right] q(\vec{r}, s).$$

A direct evaluation of this equation is not straightforward because the operators $\nabla^2$ and $\omega_{v(s)}(\vec{r})$ do not commute. In another word, these two operators cannot be diagonal at the same time. Within an error of $ds^3$, we can split these operators in the exponential to obtain an approximate solution of the propagators,

$$q(\vec{r}, s + ds) \approx \exp\left[-ds\frac{\omega_{v(s)}(\vec{r})}{2}\right] \exp[ds\sigma_{v(s)}^2 \nabla^2] \exp\left[-ds\frac{\omega_{v(s)}(\vec{r})}{2}\right] q(\vec{r}, s).$$

The advantage of this expression is that the operations of $\exp\left[-ds\frac{\omega_{v(s)}(\vec{r})}{2}\right]$ is a simple multiplication in real-space, whereas the operations of $\exp[ds\sigma_{v(s)}^2 \nabla^2]$ is a simple multiplication in the Fourier-space. An efficient numerical method to obtain solutions of the propagators can then be implemented according to the following sequence of operations,

$$q(\vec{r}, s + ds) \approx \exp\left[-ds\frac{\omega_{v(s)}(\vec{r})}{2}\right] \text{FFT}^{-1} \left\{\exp[-ds\sigma_{v(s)}^2 k^2] \text{FFT}\left(\exp\left[-ds\frac{\omega_{v(s)}(\vec{r})}{2}\right] q(\vec{r}, s)\right)\right\},$$

where FFT and FFT$^{-1}$ stand for the forward and backward (inverse) fast Fourier transform (FFT), respectively. Here the wave vectors k are defined in a discrete Fourier space according to the discretization of the real-space. The availability of highly efficient FFT algorithms allows SCFT computations being carried out in large three-dimensional boxes. The advantages of the pseudo-spectral method have made it the method of choice for the study of the self-assembly of block copolymer systems including block copolymers under confinement and multiblock copolymer melts [15].

Besides the requirement of fast and accurate methods to solve the modified diffusion equations, it has been shown that obtaining solutions of the SCFT equations depends crucially on the initialization of the mean-fields and density profiles [11]. The initialization of the fields and densities could be performed in real-space and reciprocal space. The initial structure of the system can be random or with certain symmetry. For the search and discovery of new structures, the best practice is to use a combination of different schemes to construct initial configurations. It has been shown that a combination of different initialization schemes could be used to obtain a large number of solutions for a given block copolymer system [11]. On the other hand, if the symmetry of the ordered phase is known, the initial configurations could be generated based on the symmetry of the structure either in real-space or in reciprocal space [10,15].

## 5. Application of SCFT: Complex Spherical Packing Phases of Block Copolymers

The SCFT has been applied to numerous inhomogeneous polymeric systems. A large number of ordered phases and related phase diagrams have been obtained block copolymer melts, block copolymer solution and polymer blends [15]. For completeness, in this section we present a brief summary on some recent progresses on the SCFT study of the complex spherical packing phases in block copolymers.

Within the mean-field theory, the the phase behavior of diblock copolymers is controlled by a set of parameters, $\{f, \chi N, \sigma_A, \sigma_B\}$. For the case with a small $\chi N < 10.495$, a diblock copolymer melt assumes a disordered, homogeneous phase. For larger values of $\chi N > 10.495$, the equilibrium phases of diblock copolymers changes from lamellae to gyroids to cylinders, then to spheres when the volume fraction of the A-blocks changes away from $f = \frac{1}{2}$ [1]. Among the different ordered phases of block copolymers, the self-assembled spherical phases are of particular interest because these soft materials possess the potential for advanced technological applications and their structures resemble the familiar atomic crystals. A spherical packing phase of diblock copolymers is composed of spherical domains sitting on certain lattice forming mesoscopic crystals. Prior to 2010, it was believed that the dominant stable spherical packing phase of block copolymers is the body-centered-cubic (BCC) phase, in which the spherical polymeric domains are sitting on a body-centered-cubic lattice [1]. In 2010, Bates and coworkers discovered experimentally that a non-classical and very complex spherical packing phase, the Frank-Kasper σ-phase with a giant unit cell containing 30 spheres, could become an equilibrium phase of block copolymers [21]. Further experiments showed that other complex spherical packing phases could also emerge from diblock copolymer melts [22,23]. In order to understand the origin of these complex spherical packing structures, Xie *et al.* have carried out a systematic study of the mechanisms stabilizing various complex spherical packing phases [24,25]. Specifically, these authors performed large-scale SCFT calculations for conformationally asymmetric diblock copolymers and obtained solutions of the SCFT equations corresponding to the complex spherical packing phases. These calculations used the pseudo-spectral method with designed initialization schemes. Large computational boxes with sizes up to $256 \times 256 \times 64$ were used to obtain accurate solutions for the complex spherical packing phases. Phase diagrams of diblock copolymers containing equilibrium Frank-Kasper σ-phase and A15 phase have been constructed [24,25]. Based on these SCFT results, they predicted a phase transition sequence, from BCC to Frank-Kasper σ-phase to A15 and Hexagonal Cylinders, as $f$ is increased at large $\chi N$. Most interestingly, these SCFT results showed that conformational asymmetry between the blocks is a key parameter stabilizing the complex spherical packing phases in block copolymers. This theoretical prediction has been confirmed experimentally in 2017 by Bates and coworkers using a set of designed diblock copolymer samples [26].

Further SCFT studies have demonstrated that complex spherical packing phases could be stabilized by various molecular formulations such as conformational asymmetry, topological architectures, and polydispersity distribution of macromolecules [27,28]. These theoretical studies identified the mechanisms stabilizing the different spherical phases and, more importantly, provided simpler route to engineer these complex structures. In particular, using the SCFT applied to binary blends of diblock copolymers, Liu *et al.* [27] predicted that the segregation of different polymeric species in the block copolymer blends could provide another mechanism to stabilize spherical packing phases with very different sized-spherical domains. This theoretical prediction of rich non-classical spherical packing phases in block copolymer blends provides a new and novel concept for the engineering of complex nanostructured soft matter using block copolymers. These theoretical studies of the emergence and stability of complex spherical packing phases in block copolymers clearly demonstrated the power and flexibility of the SCFT.

## 6. Conclusions

After establishing the Gibbs-Bogoliubov-Feynman (GBF) variational principle, we have shown that the mean-field theory for any statistical mechanical systems can be derived from the variational principle via a single-molecular decomposition of the total probability distribution of the system. The resulting mean-field equations can be cast in the form of a self-consistent field theory (SCFT). Application of this generic approach is demonstrated by the derivation of a mean-field density functional theory for the lattice gas model. Finally, a detailed derivation of the self-consistent field theory for inhomogeneous polymeric systems using the GBF variational principle is presented. Although the polymeric SCFT has been derived using field-theoretical methods [5,6], the variational derivation of the SCFT reveals that the nature of the mean-field approximation is the decomposition of the probability distribution into a product of one-molecular probability distributions. Compared with the field-theoretical derivation of SCFT commonly found in the literature, the variational derivation is physically more intuitive, and it does not involve complex fields.

Naturally, the derived SCFT free energy functional and SCFT equations are identical as that obtained using the field-theoretical techniques [5,6]. Numerous previous studies have demonstrated that the polymeric SCFT provides a powerful theoretical framework for the study of the phase behavior and structural properties of inhomogeneous polymeric systems such as block copolymers and polymer blends [6,15]. Although a diblock copolymer melt is used as a model system in our derivation, it should be emphasized that the variational method of deriving the SCFT is very flexible and versatile. In fact, the GBF variational principle described above is a generic approach that applies to any statistical mechanical systems. In particular, it is straightforward to extend the theoretical framework to more complex polymeric systems. For example, the SCFT framework has been extended and applied to multiblock copolymers [11,14], block copolymer blends and solutions [29], semiflexible polymers [30] and charged polymers [31].

It is interesting to note that variational approach has been used to derive density functional theory (DFT) of polymers [19,32]. In a general sense, the polymeric SCFT free energy functional could be considered as a flavour of a DFT of polymers, *albeit* within the mean-field approximation. In particular, if one could carry out the integration over the conjugate fields, the resulting free energy functional would be a functional of the polymer density alone [5]. From this perspective, the polymeric SCFT could be considered as a route to derive a DFT of polymers within the mean-field approximation.

From a mathematical point of view, the SCFT equations are in the form of a set of nonlinear and nonlocal equations. As such, obtaining analytic solutions of the SCFT equations is a formidable task. The only known exact solution of the polymeric SCFT equations is obtained for a very simple case, *e.g.* the homogeneous phase. On the other hand, approximate analytical solutions could be obtained for ordered microphases, typically formed from polymeric systems containing block copolymers, by using a number of approximate methods, such as the weak-segregation theory [33] or strong-segregation theory [34]. These approximate analytic solutions have provided very useful insight into the formation and properties of self-assembled polymeric phases. Besides analytical methods, sophisticated and accurate numerical methods to obtain solutions of the SCFT equations have been developed through the efforts of a large number of researchers [10,11,13,14,15]. The numerical solutions are exact within the numerical accuracy and they provide accurate description of the formation and structural properties of the various ordered phases self-assembled from block copolymers. Among the many numerical methods developed over the years, the pseudo-spectral method coupled with different initialization schemes provides a powerful and robust platform to explore and discover different ordered phases of polymeric systems containing block copolymers [11,15].

The variational principle used in the derivation of the mean-field theory is a generic approach that applies to any statistical mechanical systems. The essence of the approximation is the decomposition of the total probability distribution function to a product of single-molecular probability distributions. For a given molecular model, the variational free energy of the system is given in terms of the partition function of a single molecule in a mean-field. The evaluation of this single-particle partition function depends on the particularity of the molecular model. For example, if the molecule can be modeled as a classical particle, the single-particle partition function is simply obtained by carrying out the summation over the Boltzmann factor,

$$Q_\alpha(\{\omega_\alpha\}) = \frac{1}{V} \int d\vec{R} \, \exp[-\omega_\alpha(\vec{R})].$$

If the molecules are more complex, the evaluation of the single-molecular partition function is more complex. In our derivation of the polymeric SCFT, we have used Gaussian model of polymers. The advantage of this approach is that the single-polymer partition function in a field could be obtained via the solution of the modified diffusion equation. For more complex macromolecules, such as a semiflexible chain or helical wormlike chain, the calculation of the single-molecular partition function would involve solving partial differential equations in higher dimensions [30].

Another interesting class of polymeric materials is polymer-based nanocomposites, in which the molecules are flexible polymer chains and rigid nanoparticles of different sizes and shapes. In principle, the variational approach outlined in this paper can be used to derive a mean-field theory for hybrid systems such as the nanocomposites. In order to make progresses on this topic, however, two technical challenges need to be overcome. The first one is the description of the internal degrees of freedom for the nanoparticles. The single-molecular partition function for the nanoparticles must be obtained by summing over all these internal degrees of freedom. The second challenge is the development of appropriate interaction potentials between the nanoparticles and the polymers.

**Acknowledgments**

The author acknowledges the support by the Natural Science and Engineering Research Council (NSERC) of Canada.